\documentclass[prl,twocolumn,showpacs,floatfix]{revtex4}
\usepackage[dvips]{graphicx}
\begin{document}
\title{Rotational energy term in the empirical formula for the yrast energies
in even-even nuclei}
\author{Eunja \surname{Ha}}
\email{ejha@skku.edu}
\author{S. W. \surname{Hong}}
\affiliation{Department of Physics and Institute of Basic Science,
Sungkyunkwan University, Suwon 440-746, Korea}
\begin{abstract}
We show that part of the empirical formula describing the gross
features of the measured yrast energies of the natural parity even
multipole states for even-even nuclei can be related to the
rotational energy of nuclei. When the first term of the empirical
formula, $\alpha A^{-\gamma}$, is regarded as the rotational energy,
we can better understand the results of the previous analyses of the
excitation energies. We show that the values of the parameters
$\alpha$ and $\gamma$ newly obtained by considering the $\alpha
A^{-\gamma}$ term as the rotational energy of a rigid rotor are
remarkably consistent with those values extracted from the earlier
`modified' $\chi^2$ analyses, in which we use the logarithms of the
excitation energies in defining the `modified' $\chi^2$ values.
\end{abstract}


\maketitle

\pagebreak

In a series of works \cite{Ha-a, Ha-Cha, Kim, Jin-a, Jin-b}, an
empirical formula was proposed to represent the gross features of
the yrast energies $E_x$ of the natural parity even multipole states
including $2^+$, $4^+$, $6^+$, $8^+$, and $10^+$ for even-even
nuclei throughout the whole periodic table. This formula is
expressed in terms of the mass number $A$, the valence proton number
$N_p$, and the valence neutron number $N_n$ as
\begin{equation} \label{E}
E_x = \alpha A^{-\gamma} + \beta_p \exp ( - \lambda_p N_p ) +
\beta_n \exp ( - \lambda_n N_n ),
\end{equation}
where six free parameters $\alpha$, $\gamma$, $\beta_i$, and
$\lambda_i (i=p,n)$ are fixed so that the experimental excitation
energies can be fitted by Eq.\,(\ref{E}) for each multipole state
\cite{Kim}. This empirical formula has been quite successful not
only in explaining the main features of the measured excitation
energies as a function of mass number $A$ but also in reproducing
the characteristic simple patterns observed in the $N_p N_n$-plot of
the measured excitation energies \cite{Casten, Yoon}. In spite of
the success of Eq.\,(\ref{E}), however, it was not clear how this
simple formula could represent the overall features of the yrast
energies so well. Very recently, it was suggested in
Ref.\cite{Jin-b} that Eq.\,(\ref{E}) could be approximated as
\begin{equation} \label{E-mid}
E_x^{\rm mid} \approx \alpha A^{-\gamma}
\end{equation}
around the doubly mid-shell region, where $N_p$ and $N_n$ are quite
large. In this region, the two exponential terms in Eq.\,(\ref{E}),
$\beta_p \exp ( - \lambda_p N_p ) + \beta_n \exp ( - \lambda_n N_n
)$, are very small compared with the first term, $\alpha
A^{-\gamma}$. Ref.\cite{Jin-b} also suggested that the two
parameters, $\alpha$ and $\gamma$, carry the information about the
effective moment of inertia. However, it was not explained how these
parameters could be related to the effective moment of inertia.

In this Brief Report, we will first show that the first term $\alpha
A^{-\gamma}$ can be indeed expressed in terms of the effective
moment of inertia. Then we will redo some of the previous analyses
in Ref.\cite{Kim} but with the first term $\alpha A^{-\gamma}$ fixed
as the rotational energy of a rigid rotor, and compare the new
results with the previous ones.

It is well known that a nucleus near the doubly mid-shell region has
a rotational band, which consists of different total angular momenta
$J$ but shares the same intrinsic state. The energy spectrum of the
rotational band for $J^\pi$ =$2^+$, $4^+$, $6^+$, $\cdots$ with the
intrinsic angular momentum $K=0$ can be written as \cite{Bohr}
\begin{equation} \label{E-rot}
E_{\rm rot}(J^+ ) = \frac{J(J+1)\hbar^2}{2I},
\end{equation}
where $I$ is the effective moment of inertia of the nucleus. Let us
assume that a nucleus is a rigid body which has the axial symmetry
about the intrinsic 3 axis. The moment of inertia $I_{\rm rig}$ of
such a rigid body can be expressed as
\begin{equation} \label{I-rig}
I_{\rm rig}= \frac{2}{5}MR_0^2 \,(1+ \frac{\delta}{3}),
\end{equation}
where $M$ is the mass of the nucleus given by $M=Au$, $u$ being the
atomic mass unit. The distortion parameter $\delta \approx (R_3 -
R_{\bot})/R_0 $ is typically $0.2 \sim 0.3$ for nuclei with $150 \le
A \le 188$, where $R_3$, $R_{\bot}$, and $R_0$ are the radius of a
nucleus along the intrinsic $3$ axis, in the direction perpendicular
to it, and the mean radius, $R_0 =1.2 A^{1/3}$fm,
respectively\cite{Bohr}. Since the observed moments of inertia are
smaller than the moment of inertia given by the simple form in
Eq.\,(\ref{I-rig}) by roughly a factor of $2$ for nuclei around the
doubly mid-shell region with $150 \le A \le 188$\cite{Bohr}, we may
introduce a factor $k$ to take into account the difference between
the effective moment of inertia $I$ and $I_{\rm rig}$ of
Eq.\,(\ref{I-rig}) so that
\begin{equation}\label{I}
 I =kI_{\rm rig}.
\end{equation}
\begin{table*}
\centering
\begin{center}
\caption{Six parameters $\alpha$, $\gamma$, $\beta_i$, and
$\lambda_i (i=p,n)$ in the empirical formula of Eq.\,(\ref{E}) are
listed for three cases. In the upper part, four parameters $\beta_i$
and $\lambda_i$ determined by using fixed $\gamma=\gamma'=5/3$ and
$\alpha=\alpha'=\alpha_0 J (J+1)$ with $\alpha_0=65.96$ MeV are
listed: case i). In the middle part, $\alpha$, $\beta_i$, and
$\lambda_i$ determined with fixed $\gamma=1.40$ are listed: case
ii). In the lower part, the parameters previously extracted are
quoted from Table $2$ of Ref.\cite{Kim} : case iii). $N_0$ refers to
the number of data points for each multipole state. The values of
the `modified' $\chi^2$ are also listed.}

\begin{tabular} {crrrrrrrrrr}
\hline\hline
~~~$J_1^\pi$~~~~~&$\gamma$&~~~~~$\alpha$(MeV)&~~~$\alpha_0$(MeV)~&~~$\beta_p$(MeV)&
~~$\beta_n$(MeV)&~~~~~~~~$\lambda_p$&~~~~~~~~$\lambda_n$&~~~~~~~~~$\chi^2$&~~~~~~$N_0$&\\
\hline
$2_1^+$ & 1.67 & 395.76  &65.96  & 0.79 & 1.09 & 0.42 & 0.29 & 0.157 & 557 &\\
$4_1^+$ & 1.67 & 1319.20 &65.96  & 1.12 & 1.54 & 0.34 & 0.24 & 0.094 & 430 &\\
$6_1^+$ & 1.67 & 2770.32 &65.96  & 1.31 & 1.46 & 0.32 & 0.18 & 0.086 & 375 &\\
$8_1^+$ & 1.67 & 4749.12 &65.96  & 1.27 & 1.34 & 0.26 & 0.17 & 0.060 & 309 &\\
$10_1^+$& 1.67 & 7255.60 &65.96  & 1.30 & 1.46 & 0.23 & 0.18 & 0.040 & 265 &\\
\hline
$2_1^+$ & 1.40 & 89.89 &14.98  & 0.82 & 1.15 & 0.41 & 0.28 & 0.126 & 557 &\\
$4_1^+$ & 1.40 & 297.87 &14.89 & 1.20 & 1.67 & 0.33 & 0.23 & 0.071 & 430 &\\
$6_1^+$ & 1.40 & 654.71 &15.59 & 1.40 & 1.64 &0.31 & 0.18  & 0.069 & 375 &\\
$8_1^+$ & 1.40 & 1155.90&16.05 & 1.34 & 1.50 & 0.26 & 0.15 & 0.053 & 309 &\\
$10_1^+$& 1.40 & 1702.79 &15.48& 1.34 & 1.64 & 0.22 & 0.15 & 0.034 & 265 &\\
\hline
$2_1^+$ & 1.34 & 68.37 &11.40  & 0.83 & 1.17 & 0.42 & 0.28 & 0.126 & 557 &\\
$4_1^+$ & 1.38 & 268.04 &13.40 & 1.21 & 1.68 & 0.33 & 0.23 & 0.071 & 430 &\\
$6_1^+$ & 1.38 & 598.17 &14.24 & 1.40 & 1.64 & 0.31 & 0.18 & 0.069 & 375 &\\
$8_1^+$ & 1.45 & 1438.59 &19.98& 1.34 & 1.50 & 0.26 & 0.15 & 0.053 & 309 &\\
$10_1^+$& 1.47 & 2316.85 &21.06& 1.36 & 1.65 & 0.21 & 0.14 & 0.034 & 265 &\\
\hline\hline
\end{tabular}
\label{tab-para}
\end{center}
\end{table*}
By inserting Eq.\,(\ref{I}) into Eq.\,(\ref{E-rot}), we immediately
get
\begin{equation}\label{E-rot-mid}
E_{\rm rot}(J^+)=\alpha' \,A^{-\gamma'}
\end{equation}
where
\begin{eqnarray}\label{param}
\alpha'&=&\alpha_0 J(J+1)~~{\rm and}~~ \gamma'= 5/3 ~~~~~~~~~~ \\
{\rm with}~~~~~~~~~~~~~~~~~~&&\nonumber \\
\alpha_0 &=&\frac{5\hbar^2}{4uk1.2^2 (1+\frac{\delta}{3})}.\nonumber
\end{eqnarray}
\begin{figure}[h]
\centering
\includegraphics[width=8.5cm,angle=0]{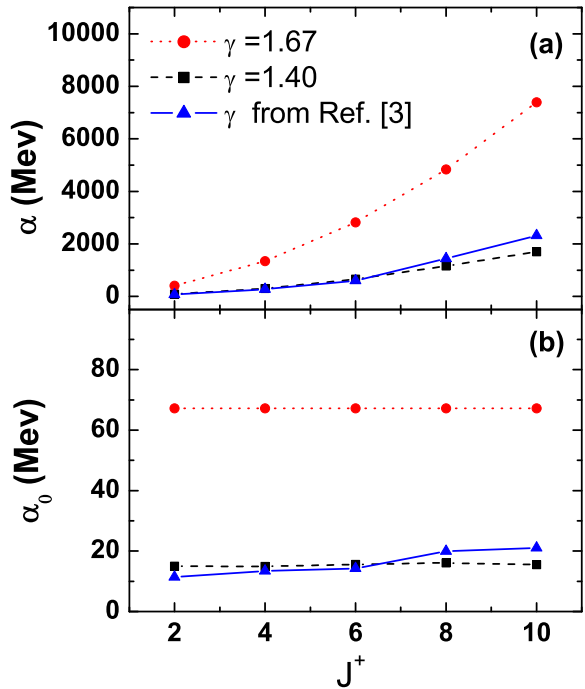}
\caption{(Color online) The values of $\alpha$ and $\alpha_0$ in
Table \ref{tab-para} are plotted for the following three case: i)
$\gamma=1.67$ (circles), ii) $\gamma=1.40$ (squares), and iii)
$\gamma$ from Ref.\cite{Kim} (triangles).} \label{fig-alpha}
\end{figure}
\begin{figure}[t]
\centering
\includegraphics[width=8.5cm,angle=0]{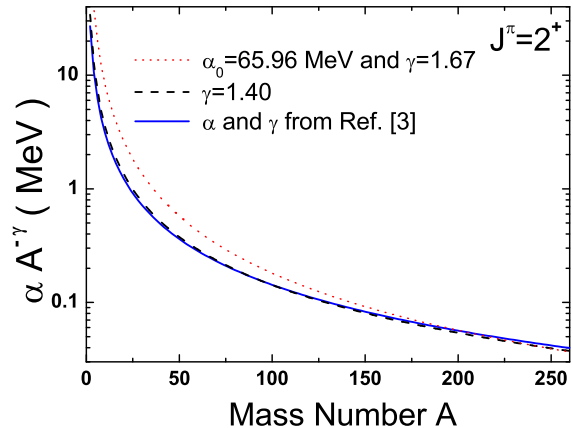}
\caption{(Color online) The first term of
Eq.\,(\ref{E}) for the $2_1^+$ states in even-even nuclei are
plotted against the mass number $A$. The solid curve represents the
values of $\alpha A^{-\gamma}$ calculated with $\alpha$ and $\gamma$
from Ref.\cite{Kim}. The dashed curve denotes the values of $\alpha
A^{-\gamma}$ with $\gamma = 1.40$ and $\alpha$
determined by the $\chi^2$ fitting. The dotted curve denotes
$\alpha A^{-\gamma}$ when $\alpha_0$ and $\gamma$ are
fixed as $65.96$ MeV and $1.67$, respectively.} \label{fig-first}
\end{figure}
\begin{figure*}
\centering
\includegraphics[width=18.0cm,angle=0]{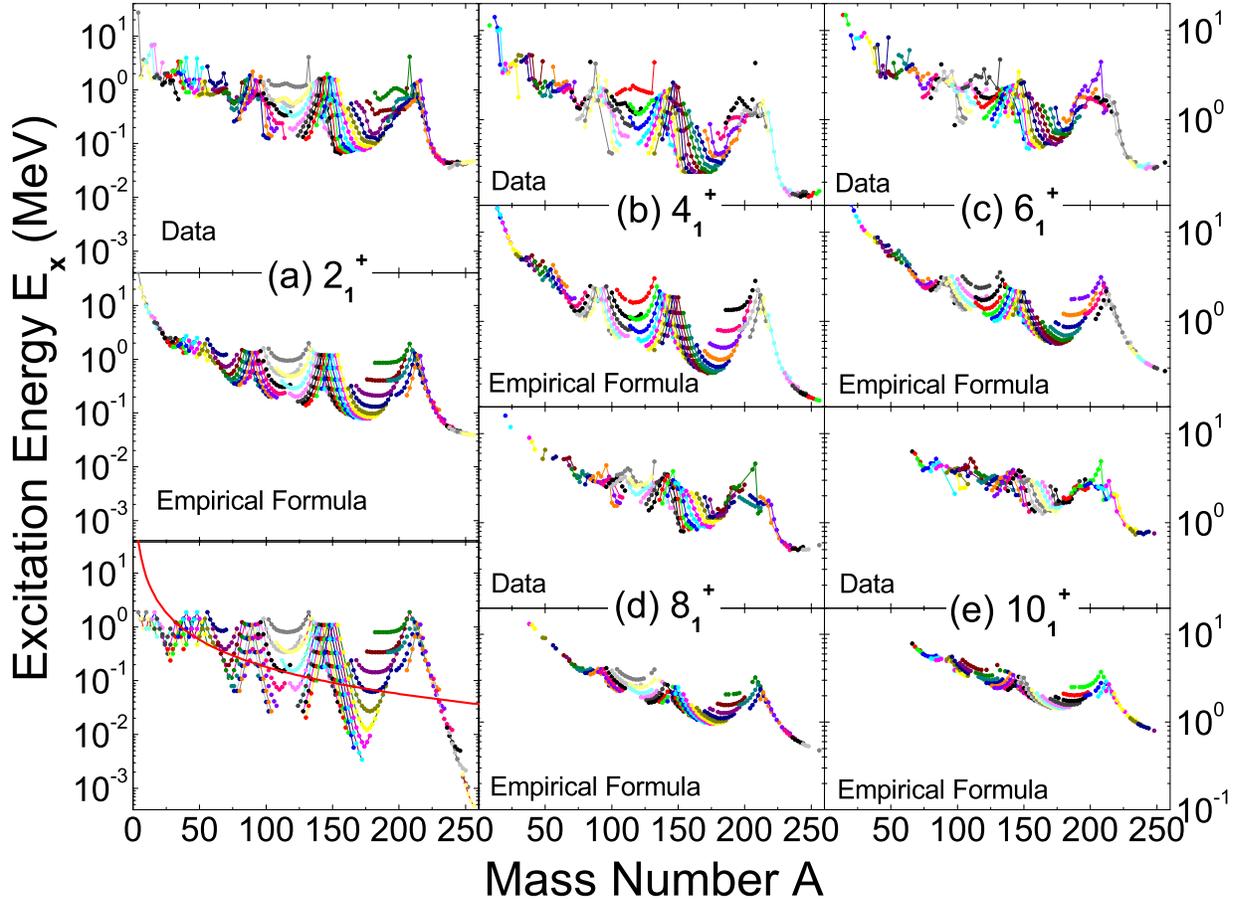}
\caption{(Color online) The measured excitation energies of the
$2_1^+$, $4_1^+$, $6_1^+$, $8_1^+$, and $10_1^+$ states in even-even
nuclei are compared with our results. The measured data are
connected by the solid lines along the isotopic chains. The upper
parts of (a)-(e) show the measured energies \cite{Raman,Firestone},
while the lower parts show the calculated energies by using the
parameter sets in the upper part of Table \ref{tab-para}. The solid
curve and the circles in the bottom part of (a) show the
contribution to the calculated energies of the $2_1^+$ states from
the first term and two exponential terms. } \label{fig-all}
\end{figure*}
\begin{figure}
\centering
\includegraphics[width=8.5cm,angle=0]{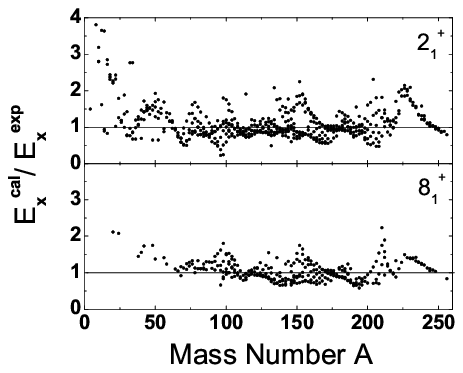}
\caption{The ratios of the calculated excitation energies to the
measured energies for the $2_1^+$ and $8_1^+$, $E_x^{{\rm cal}} /
E_x^{{\rm exp}}$, are plotted against the mass number A.}
\label{fig-com}
\end{figure}
Eq.\,(\ref{E-rot-mid}) shows that the first term $\alpha
A^{-\gamma}$ of the empirical formula in Eq.\,(\ref{E}) can be
derived and identified as the rotational energy of a deformed
nucleus. However, this interpretation of the first term $\alpha
A^{-\gamma}$ is valid only for nuclei near the doubly mid-shell
region, but not for closed shell nuclei or for light nuclei with $A
\lesssim 30$, most of which do not have rotational bands. For closed
shell nuclei, on the other hand, the first term $\alpha A^{-\gamma}$
is negligible compared to the other two exponential terms, which
will be shown later in Fig.3. Thus if we accept this interpretation
to this extent, Eq.\,(\ref{E-rot-mid}) shows that $\gamma$ in
Eq.\,(\ref{E}) can be identified as a constant $\gamma'$ close to
$1.67$ for all of the even multipole states. Note that the values of
$\gamma$ in Eq.\,(\ref{E}) extracted earlier and shown in Table $2$
of Ref.\cite{Kim} range from $1.34$ to $1.47$. These previous values
of $\gamma$ are listed again in the lower part of Table
\ref{tab-para}. The average of these previous $\gamma$ values is
$1.40$ which is remarkably close to $\gamma'=5/3$, in spite of the
fact that the values of $\gamma$ were extracted earlier without any
constraints on $\gamma$ or consideration presented in
Eqs.\,(\ref{I})$\sim$(\ref{param}).

Let us repeat the $\chi^2$ analyses as was done in Ref.\,\cite{Kim}
but with a constant average value of $\gamma=1.40$. In the fitting
procedure we use as in Ref.\,\cite{Kim} the logarithm of the
excitation energies $E_x^{\rm cal}(i)$ and $E_x^{\rm exp}(i)$ since
the excitation energies are spread over a wide range. This
`modified' $\chi^2$ analysis is different from the conventional
$\chi^2$ analysis in that the small values of the excitation
energies are emphasized. By defining $R_{\rm E} (i) = \rm log
\left[E_x^{\rm cal} (i)\right]-log \left[ E_x^{\rm exp} (i)\right]$,
the five parameters $\alpha$, $\beta_i$ and $\lambda_i$ are fixed to
minimize $\chi^2 = { 1 \over {N_0}} \sum_{i=1}^{N_0} \left[ R_{\rm
E} (i) \right]^2$, where $N_0$ is the total number of the data
points considered for the corresponding multipole state. $\alpha$,
$\beta_i$, and $\lambda_i$ extracted in this way are given in the
middle part of Table \ref{tab-para}. The values of $\alpha$ and
$\alpha_0 =\alpha / J(J+1)$ determined with $\gamma=1.40$ are
somewhat different from those of $\alpha$ and $\alpha_0$ in the
lower part of Table \ref{tab-para}. On the other hand, the values of
$\beta_i$ and $\lambda_i$ in the middle part of Table \ref{tab-para}
are still very close to those values in the lower part for all the
multipole states. This shows that the two exponential terms $\beta_p
\exp ( - \lambda_p N_p ) + \beta_n \exp ( - \lambda_n N_n )$ remain
rather robust even if $\alpha$ and $\gamma$ change and thus that the
two exponential terms can be well separated from the rotational
energy term.

Let us now fix $\alpha$ and $\gamma$
as given by Eq.\,(\ref{param}) for all $J$ and redo the `modified'
$\chi^2$ analyses for $2^+$ to $10^+$ states. In the analyses we use
$k=1/2$ and the distortion parameter $\delta=0.3$ determined for
${\rm ^{174}Yb} \,(N_p=12, \, N_n=22)$ which is a nucleus in the
doubly mid-shell region \cite{Bohr}. Then the value of $\alpha_0$
becomes $65.96$ MeV. In Table \ref{tab-para} we show that the
`modified' $\chi^2$ values obtained by using these fixed values of
$\alpha'$ and $\gamma'$ of Eq.\,(\ref{param}) turn out to be larger
than those previously obtained \cite{Kim} by $25\%$ (for $2_1^+$),
$32\%$ (for $4_1^+$), $25\%$ (for $6_1^+$), $13\%$ (for $8_1^+$),
and $18\%$ (for $10_1^+$). Nevertheless, the excitation energies
calculated by using fixed $\alpha_0 =65.96$ MeV and $\gamma'=5/3$
appear almost identical to those obtained earlier in \cite{Kim}.

To compare the different values of $\alpha$ listed in Table
\ref{tab-para}, we plot in Fig. \ref{fig-alpha}
the values of $\alpha$ and $\alpha_0$ against $J$ for three cases: i)
$\gamma=\gamma'=1.67$ (circles), ii) $\gamma=1.40$ (squares), and
iii) $\gamma$ from Ref.\cite{Kim} (triangles). $\alpha$'s for all
three cases in Fig. \ref{fig-alpha}(a) change as a quadratic
function of $J$. Thus we plot in Fig. \ref{fig-alpha}(b) $\alpha_0
=\alpha / J(J+1)$. Figure \ref{fig-alpha}(b) shows that $\alpha_0$'s
are indeed constant for all three cases. It is remarkable to see
that the values of $\alpha_0$ (triangles) obtained from those of
$\alpha$ extracted previously in Ref.\cite{Kim} turn out to be also
almost constant. Figure \ref{fig-alpha}
shows that the values of $\alpha$ for the cases
ii) and iii) are very close to each other but are smaller than those
for the case i) roughly by a factor $4$. This factor is to
compensate for the changes in $\alpha A^{-\gamma}$ due to the
changes in $\gamma$ from about $1.40$ to $1.67$. Note that the ratio
$A^{-1.40}/A^{-1.67}=A^{0.27}$ is about $4$ for $^{174}Yb$.

In Fig.\,\ref{fig-first} the value of $\alpha A^{-\gamma}$ term with
$\alpha$ and $\gamma$ as given in Ref.\cite{Kim} is plotted by the
solid curve, and that with $\gamma=1.40$ is plotted by the dashed
curve. They agree very well. The values of $E_{\rm rot} (J^+ )$ of
Eq.\,(\ref{E-rot-mid})
with $\alpha'$ and $\gamma'$ given in Eq.\,(\ref{param}) are plotted
by the dotted curve, which also agree with the other curves quite
well for larger $A$. This shows that the values of $\alpha$
compensate for the changes in the values of $\gamma$ so that $\alpha
A^{-\gamma}$ remains more or less the same. Note, however, that the
dotted curve deviates from the other two curves
for small values of $A$. It is because we use the logarithm of
energies rather than the energies themselves for the 'modified'
$\chi^2$ calculations. From the $\chi^2$ analyses with $\alpha_0
=65.96$ MeV and $\gamma=1.67$ we obtain new parameters $\beta_i$ and
$\lambda_i$ listed in the upper part of Table \ref{tab-para}. The
new parameters of $\beta_i$ and $\lambda_i$ are still very close to
those values in the middle and lower parts. As mentioned earlier,
the parameters of $\beta_i$ and $\lambda_i$ are almost the same for
three cases of i) $\sim$ iii), implying that the two exponential
terms are rather independent of the rotational energy term $\alpha
A^{-\gamma}$.

In Fig.\,\ref{fig-all} we show the yrast energies for each state of
$2_1^+$(a), $4_1^+$(b), $6_1^+$(c), $8_1^+$(d), and $10_1^+$(e)
against the mass number $A$. For each $J^{\pi}$ state the
experimental yrast energies are plotted in the upper part and are
compared with the calculated values in the lower part obtained by
using the parameters given in the upper part of Table
\ref{tab-para}. The measured excitation energies of the first $2^+$
are quoted from the compilation of Raman {\it et al}.\cite{Raman}
and those of the first $4^+ \sim 10^+$ are extracted from
Ref.\,\cite{Firestone}. The data points are connected by the solid
lines along the isotopic chains. Figure\,\ref{fig-all} shows that
Eq.\,(\ref{E}) with fixed $\alpha'$ and $\gamma'$ in
Eq.\,(\ref{param}) can reproduce the overall trends of the measured
yrast energies for all of the states. To show the discrepancies
between the experimental energies and the calculated energies, we
plot the ratios $E_x^{{\rm cal}} / E_x^{{\rm exp}}$ against the mass
number $A$ in Fig.\,\ref{fig-com}. 
(We plot the ratios only for the $2_1^+$ and $8_1^+$ states,
because the ratios for the other states 
are quite similar.) 
The ratios are close to unity for
nuclei near the doubly mid-shell region. For nuclei in the closed
shell region where our interpretation of $\alpha A^{-\gamma}$ is not
valid, the ratios can be as big as $2$. For light nuclei to which
our interpretation is not applicable $E_x^{{\rm cal}}$ is often much
larger than $E_x^{{\rm exp}}$. This is because the `modified'
$\chi^2$ analyses put more emphasis on the smaller values of the
energies and thus $\alpha A^{-\gamma}$ with $\gamma=1.67$ (dotted
curve in Fig.\,\ref{fig-first}) overshoots $\alpha A^{-\gamma}$ with
$\gamma=1.40$ (dashed curve in Fig.\,\ref{fig-first}) for small
values of $A$.
Finally we show in the bottom part of Fig.3(a)
the separation of the calculated energies into the first term
$\alpha A^{-\gamma}$ and the two exponential terms. The first term
$\alpha A^{-\gamma}$ (the solid curve) is the major contribution in
the yrast energies in the doubly mid-shell region while it is
negligible compared to the two exponential terms (circles) in the
closed shell region.

In summary, we have shown that the term $\alpha A^{-\gamma}$
can be obtained by considering the moment of
inertia of a deformed nucleus. The yrast energies calculated with
constant $\alpha'$ and $\gamma'$ can describe the main features of
the data. It is remarkable that the
values of $\alpha A^{-\gamma}$ extracted earlier in Ref.\,\cite{Kim}
agree well with the values of $\alpha' A^{-\gamma'}$ obtained from
the rotor model and that the previous values of $\alpha$ divided by
$J(J+1)$ are almost constant as expected from the rotor model. It is
also seen that the parameters $\beta_i$ and $\lambda_i$ newly
extracted with constant $\alpha'$ and $\gamma'$ in this work are
very consistent with those parameters previously obtained in
Ref.\,\cite{Kim}. It shows that the empirical formula
can be well separated into the
rotational energy term $\alpha A^{-\gamma}$ and the two exponential
terms which are thought to be related to the shell effect.
\begin{acknowledgments}
The authors would like to thank Prof. D. Cha for the helpful
discussions. This work was supported in part by Faculty Research
Fund of Sungkyunwan University 2007,
the KRF Grant funded by the Korean Government(MOEHRD)
(KRF-2006-312-C00506)
and the KOSEF grant funded by the Korean Government (MEST)
(No. M20608520001-08B0852-00110).
\end{acknowledgments}

\end{document}